\documentclass[11pt,twocolumn]{article}
\usepackage[utf8]{inputenc}
\usepackage{lmodern}
\usepackage{amsmath}
\usepackage{amsfonts}
\usepackage{array}
\usepackage{amssymb}
\usepackage[version=4]{mhchem}
\usepackage{stmaryrd}
\usepackage{bbold}
\usepackage{caption}
\usepackage[export]{adjustbox}
\usepackage[margin=0.75in]{geometry}
\usepackage{cancel}
\usepackage{subfig}
\usepackage{xcolor}
\usepackage{framed}
\usepackage{floatrow}
\usepackage{listings}
\usepackage{enumitem}
\usepackage{graphicx}
\usepackage{ulem}
\usepackage{mathtools}
\usepackage{amsopn}
\usepackage[colorlinks=true,linkcolor=black,citecolor=blue,urlcolor=blue]{hyperref}
\usepackage{mathtools}
\usepackage{algorithm}
\usepackage{algpseudocode}
\usepackage{listings}
\usepackage{placeins}
\usepackage{amsthm}
\usepackage{longtable}
\usepackage{booktabs}
\usepackage{array}
\usepackage{titlesec}
\usepackage{etoolbox}
\usepackage{stfloats}
\usepackage[backend=biber,style=numeric,giveninits=true,minnames=1]{biblatex}
\mathtoolsset{showonlyrefs=true}
\setlistdepth{6}
\renewlist{itemize}{itemize}{6}

\titlespacing*{\section}
{0pt}{1.25ex}{0.8ex}
\titlespacing*{\subsection}
{0pt}{1.0ex}{0.5ex}
\titlespacing*{\paragraph}
{0pt}{0.75ex}{0.5ex}
\renewenvironment{abstract}
  {\small
   \begin{center}
   \bfseries\abstractname
   \end{center}
   \vspace{-0.25em}
   \quotation}
  {\endquotation}
\patchcmd{\abstract}
  {\quotation}
  {\quotation\vspace{-1.2em}}
  {}{}
\newcommand{\ital}[1]{\textit{#1}}
\newcommand{\boldt}[1]{\textbf{#1}}

\newcommand{\Cov}{\operatorname{Cov}}

\newcommand{\indentation}{\\ $\phantom{}\qquad\ $}

\newtheoremstyle{dotstyle}
  {1.5em}       
  {1em}         
  {\itshape}    
  {}            
  {\bfseries}   
  {.}           
  {0.5em}       
  {}            
\theoremstyle{dotstyle}

\newcounter{maincounter}

\newtheorem{assumption}[maincounter]{\boldt{Assumption}}

\newtheorem{claim}[maincounter]{\boldt{Claim}}

\newtheorem{corollary}[maincounter]{\boldt{Corollary}}

\newtheorem{counterexample}[maincounter]{\boldt{Counterxample}}

\newtheorem{definition}[maincounter]{\boldt{Definition}}

\newtheorem{example}[maincounter]{\boldt{Example}}

\newtheorem{fact}[maincounter]{\boldt{Fact}}

\newtheorem{lemma}[maincounter]{\boldt{Lemma}}

\newtheorem{proposition}[maincounter]{\boldt{Proposition}}

\newtheorem{remark}[maincounter]{\boldt{Remark}}
\newcommand{\Remark}[2]{
    \begin{remark}
        #1
        \label{remark: #2}
    \end{remark}
}

\newtheorem{theorem}[maincounter]{\boldt{Theorem}}

\begin{document}

    \title{From Cointegration to Out-of-Sample Failure:\\ A Pairs-Trading Case Study on PEP/KO}
    \author{Graziano, Davide}
    \date{}
    \maketitle

    \begin{abstract}

    This paper examines whether a cointegration-based pairs trading strategy between PepsiCo and The Coca-Cola Company is statistically robust and economically exploitable. 
    We first test for cointegration and estimate the spread's mean-reversion dynamics over 2013--2018, then hold these statistical parameters fixed and optimise a threshold-based trading strategy in-sample over 2018--2023.
    Robustness is assessed through transaction-cost and parameter sensitivity tests, walk-forward validation, and Adjusted and Deflated Sharpe Ratios. 
    The strategy is then evaluated out-of-sample from 2023 to the present, including an analysis of time-varying hedge ratios using rolling OLS and a Kalman filter. 
    The results show that weakening mean-reversion dynamics in the spread undermine the effectiveness of the strategy out-of-sample.

\end{abstract}

    \section{Introduction}
\label{section: introduction}

Pairs trading is a market-neutral strategy that seeks to profit from temporary mispricings between two historically related assets. 
A common statistical foundation for the strategy is \ital{cointegration}: although two asset prices may each be non-stationary, a linear combination of them can be stationary. 
This long-run trend suggests that deviations from their equilibrium relationship may eventually reverse. 
If the relationship is sufficiently stable, such deviations can potentially be exploited through a systematic trading strategy.
\indentation
An extensive body of literature examines the profitability of pairs trading. 
In their seminal case study, \textcite{Gatev_Goetzmann_Rouwenhorst_Pairs_Trading} show that pairs trading generates significant excess returns by exploiting temporary mispricing between economically related securities, with the results remaining robust after accounting for transaction costs. 
Subsequent studies formalised pair selection using the cointegration framework, identifying securities with stable long-run relationships \cite{Vidyamurthy_Pairs_Trading,Elliot_VanDerHoek_Malcolm_Pairs_Trading}. 
However, later evidence suggests that pairs-trading profitability may be unstable over time and sensitive to transaction costs and changes in market conditions \cite{Do_Faff_Pairs_Trading_Profitability}.
\indentation
We examine this issue using PepsiCo (PEP) and The Coca-Cola Company (KO), a commonly studied economically related pair \cite{Zhang_PEP_KO_Pairs_Trading}. 
Although both firms operate in the global beverage industry and exhibit similar price dynamics, existing evidence finds that their cointegrating relationship may be weak and period-dependent \cite{Cui_PEP_KO_Cointegration}. 
We therefore study whether the PEP--KO relationship remains stable and economically exploitable from 2013 to the present, after accounting for transaction costs, model-selection bias, and potential regime changes.
\indentation
The analysis proceeds in four stages.
The first step is classical: we examine the stationarity properties of the individual log-price and return series using the 2013-2018 daily price data using the Augmented Dickey--Fuller test, and then apply the Engle--Granger procedure to test for cointegration. 
The resulting spread is further characterised as an $AR(1)$ model, from which we estimate its mean-reversion speed and half-life.
Second, we exploit these findings to justify a rule-based trading policy governed by entry and exit thresholds on the standardised spread, and we optimise this policy in-sample over a 2018--2023 training window. 
Third, we examine the robustness of the resulting strategy through sensitivity tests on transaction costs, risk budgets, and volatility-window lengths, as well as through walk-forward validation. 
We also investigate the effect of the COVID-19 volatility shock and apply both the Adjusted Sharpe Ratio and Deflated Sharpe Ratio to account for non-Gaussian returns and the selection bias associated with searching across multiple candidate policies.
Finally, the selected strategy is evaluated out-of-sample (OOS) from 2023 to the present. 
We assess whether the hedge ratio estimated during the 2013-2018 period remains appropriate over time and consider two approaches that allow it to evolve dynamically: rolling OLS and a Kalman filter.
\indentation 
Throughout, particular care is taken to avoid look-ahead bias and to distinguish between statistical parameters, estimated using data from the 2013--2018 window, and trading parameters, which are optimised in-sample during the 2018--2023 training window and subsequently evaluated (but not re-optimised) OOS.
Moreover, performance statistics such as the Sharpe ratio are carefully interpreted in light of the number of trades and the number of policies searched over.

\Remark{
    Throughout, prices are retrieved as \ital{adjusted} daily closing prices, i.e.\ corrected for dividends and stock splits. 
    Unadjusted prices exhibit a drop on each ex-dividend date that is unrelated to the underlying price dynamics which we seek to model. 
    Left uncorrected, such drops would appear as spurious negative jumps in the return series $r_t$, potentially triggering false trading signals in Section~\textnormal{\ref{section: trading policy}}.
}{adjusted prices}

    \section{Stationarity and Cointegration}
\label{section: statistical analysis}

The statistical analysis is based on the prices of PEP and KO over the period 2013--2018.
Let $P_t$ denote the closing price of a given company at time $t$, and denote its log-price and log-return by $X_t := \log P_t$ and $r_t := X_t - X_{t-1}$, respectively.
Over the sample period, the annualised volatility of PEP and KO log-returns is $13.09\%$ and $13.80\%$ respectively, consistent with two large, mature and low-growth companies.
Their daily log-returns exhibit a sample correlation of $\hat\rho_{\text{PEP,KO}} = 0.66$.
Although the two return series exhibit a strong contemporaneous association, correlation alone does not establish a long-run relationship between the underlying price processes. 
Since correlations between non-stationary series may be spurious, determining the order of integration of each price series is a necessary first step before testing for cointegration.
\indentation
We model the log-price process as a first-order autoregressive process, $AR(1)$,
\begin{equation}
    X_t = \alpha + \rho X_{t-1} + \varepsilon_t, 
    \qquad \varepsilon_t \overset{\mathrm{i.i.d.}}{\sim} \mathcal{N}(0,\sigma_\varepsilon^2),
    \label{eq: AR model of order 1}
\end{equation}
where we assume $\Cov(X_0,\varepsilon_t)=0$ for all $t$.
The qualitative behaviour of the process is governed by $\rho$: the process is weakly stationary if $|\rho|<1$, has a unit root if $\rho=1$, and is explosive if $|\rho|>1$.

\subsection{Unit Root Testing}
Before conducting a formal unit-root test, a qualitative inspection of the price dynamics rules out, to a reasonable approximation, both explosive and alternating behaviour. 
We therefore restrict attention to the case $\rho\in(-1,1]$.
Within this restricted parameter space, testing $H_0:\rho=1$ against $H_1:\rho<1$ constitutes a binary hypothesis test, which we carry out using the Augmented Dickey--Fuller (ADF) test \cite{Dickey_Fuller_ADF_Test}. 
The lag order is selected according to the Akaike Information Criterion (AIC), which trades off goodness of fit against model complexity.
Applied to the log-price series, the ADF test fails to reject the unit-root null for either company at the $5\%$ level. 
Conversely, applying the same test to the log-return series $r_t=\Delta X_t$ results in a strong rejection of the unit-root null for both companies at the $5\%$ level (see Table~\ref{tab:adf-results} for both tests). 
Taken together, these results indicate that both log-price series are $I(1)$: they are non-stationary in levels but stationary after first differencing.

\begin{table}[h]
\centering
\begin{tabular}{lccc}
\toprule
Series & ADF statistic & $p$-value & $5\%$ CV \\
\midrule
$\{X_t^{PEP}\}$ & $-1.545$ & $0.511$ & $-2.864$ \\
$\{X_t^{KO}\}$  & $-1.663$ & $0.450$ & $-2.864$ \\
$\{r_t^{PEP}\}$ & $-36.50$ & $<0.001$ & $-2.864$ \\
$\{r_t^{KO}\}$  & $-36.62$ & $<0.001$ & $-2.864$ \\
\bottomrule
\end{tabular}
\caption{ADF test results for log-prices and log-returns, 2013--2018.}
\label{tab:adf-results}
\end{table}

\subsection{Cointegration and Spread Dynamics}

Since both log-price series are $I(1)$, we next investigate whether they are cointegrated in the sense of \textcite{Engle_Granger_Test}. 
In particular, cointegration asks if there exist $\alpha,\beta\in\mathbb{R}$ such that the linear combination
\begin{equation}
    u_t := X_t^{\text{PEP}}-\beta X_t^{\text{KO}}-\alpha
    \label{eq: spread}
\end{equation}
is stationary. 
We estimate $(\alpha,\beta)$ by OLS regression of $X_t^{\text{PEP}}$ on $X_t^{\text{KO}}$, obtaining $\hat\alpha=-1.409$ and $\hat\beta=1.662$.
We then apply the Engle--Granger test to the estimated residuals $\hat u_t = X_t^{\text{PEP}}-\hat\beta X_t^{\text{KO}}-\hat\alpha$ -- we use the adjusted critical values to account for the fact that the cointegrating relationship is estimated from the data. 
The adjusted test rejects the null hypothesis of no cointegration at the $5\%$ significance level ($p$-value 0.033), providing evidence of a long-run equilibrium relationship between the PEP and KO log-prices over the sample period.
\indentation
Having established the stationarity of $\hat u_t$, we characterise its dynamics by fitting the $AR(1)$ model
\begin{equation}
    \hat u_t = a + \rho_u \hat u_{t-1} + \eta_t,
\end{equation}
which yields $\hat\rho_u=0.9785$ and an implied equilibrium level
$$
    \hat m=\frac{\hat a}{1-\hat\rho_u}= 0.0037,
$$
which is effectively zero. 
The corresponding mean-reversion speed $\hat\kappa:=1-\hat\rho_u= 0.0215$ implies a half-life of mean reversion of approximately $31.9$ trading days. 
Together, the estimated cointegrating relationship and the dynamics of the stationary spread provide the statistical basis for the entry, exit, and volatility-window parameters of the trading policy developed in the following section.

    \section{Trading Policy}
\label{section: trading policy}

We translate the estimated spread dynamics into a rule-based trading policy comprising entry and exit thresholds, position sizing, and an explicit treatment of transaction costs.

\paragraph{Entry and exit.} 
We standardise the estimated spread $\hat u_t$ into a $z$-score,
\begin{equation}
    z_t := \frac{\hat u_t - \hat m}{\hat\sigma_u},
    \label{eq: z-score}
\end{equation}
where $\hat m$ and $\hat\sigma_u$ are, respectively, the equilibrium level and sample standard deviation of the spread estimated previously.
Hence, $z_t$ measures how many standard deviations the spread currently lies from its long-run equilibrium.
A large positive $z_t$ indicates an unusually high spread: under the mean-reversion assumption, it is expected to decline toward equilibrium; a large negative $z_t$ indicates an unusually low spread and is expected to rise.
The trading policy is therefore governed by $\theta:=(z_{\text{entry}},z_{\text{exit}})$, with $z_{\text{entry}}>0$ and $0\le z_{\text{exit}}<z_{\text{entry}}$. 
When $z_t$ crosses above $z_{\text{entry}}$, the strategy enters a \ital{short-spread} position (short one unit of PEP, long $\hat\beta$ units of KO), which profits from a subsequent decline in the spread; when $z_t$ crosses below $-z_{\text{entry}}$, it enters the opposite \ital{long-spread} position.
In both cases, the position is closed once $|z_t|$ falls below $z_{\text{exit}}$, at which point the spread has reverted sufficiently close to equilibrium that any further reversion no longer justifies holding the position.

\paragraph{Position sizing.} 
Sizing is fixed ex ante by a risk-budgeting rule and is not an optimisation parameter. 
Let $C_t$ denote total portfolio equity at time $t$, inclusive of both realised and unrealised profits and losses, with initial capital $C_0=\$100{,}000$, and let $w$ denote the fixed fraction of equity allocated as nominal risk budget at each time step, $w\cdot C_t$.
The position is scaled inversely to the recent volatility $\hat\sigma_{\Delta z}^L$ of the spread's daily changes $\Delta z_t = z_t-z_{t-1}$, estimated over a trailing window of $L$ days up to $t-1$ to avoid look-ahead. 
Because $z_t$ is standardised, $\hat\sigma_{\Delta z}^L$ does not capture the magnitude of the spread's movements. 
The additional factor $\hat\sigma_u$ is therefore included to restore the scale of the underlying spread's volatility.
The resulting dollar notional exposure is
\begin{equation}
    N_t = C_t\min\left(\frac{w}{\hat\sigma_u\,\hat\sigma_{\Delta z}^L},\,
    N_{\max}\right),
    \label{eq: sizing rule}
\end{equation}
where $N_{\max}$ caps the maximum fraction of current equity for the position. 
For a long position in the spread, the corresponding share counts are therefore $S_t^{\text{PEP}} = N_t/P_t^{\text{PEP}}$ and $S_t^{\text{KO}} = -\hat\beta\, N_t/P_t^{\text{KO}}$, with signs reversed for a short position.

\paragraph{Transaction costs.} 
A transaction cost $c$, expressed as a fraction of traded notional, is applied once on entry and once on exit of each leg. 
The costs are deducted directly from $C_t$, so that they enter the objective function (see below) directly rather than being a post-hoc adjustment.
This is especially important because narrower entry/exit bands capture smaller reversions but trigger more frequent round-trip trades, so a cost-blind criterion can favour more active policies.

\paragraph{Optimisation objective.} 
Let 
\begin{equation*}
    \Pi_t(\theta) = \frac{C_t(\theta)-C_{t-1}(\theta)}{C_{t-1}(\theta)}
\end{equation*}
denote the daily return on capital for a given policy $\theta$.
The optimal policy $\theta^\star$ is selected by maximising the annualised net Sharpe ratio
\begin{equation}
    \theta^\star \in \operatorname*{arg\,max}_{\theta\in\Theta}
    J(\theta), \qquad J(\theta) = \frac{\hat\mu_\Pi(\theta)}
    {\hat\sigma_\Pi(\theta)}\sqrt{252},
    \label{eq: annualised net Sharpe ratio - objective function}
\end{equation}
over a predefined grid of candidate policies $\Theta$. 
Raw net P\&L is deliberately avoided as an objective, since it may reward highly levered or infrequent profitable trades without penalising risk.

\begin{figure}[b]
    \centering
    \includegraphics[width=\textwidth]{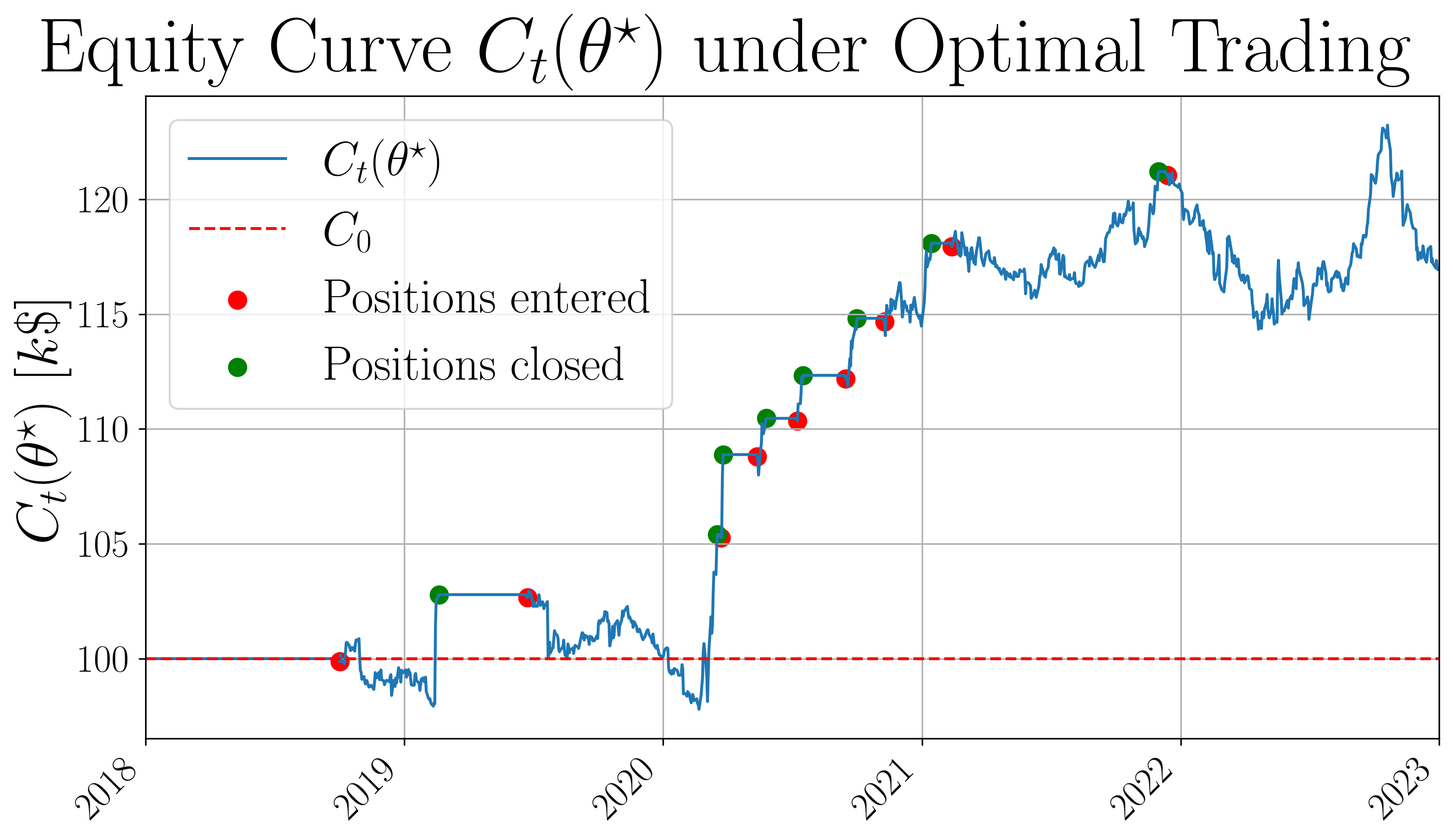}
    \caption{Equity curve under optimal trading during the 2018--2023 training window.}
    \label{fig: optimal equity curve 2018-2023}
\end{figure}

\subsection{Policy Optimisation Results}
The policy parameters other than $\theta$ are fixed ex ante, with a risk budget $w=0.5\%$, a maximum notional exposure $N_{\max}=25\%$, a transaction cost $c=0.20\%$, and a volatility-estimation window $L=64$ trading days (approximately twice the mean-reversion half-life estimated in Section~\ref{section: statistical analysis}).
Optimising $J$ over the threshold grid, using only the 2018--2023 training window, selects
\begin{equation}
    \theta^\star = (z_{\text{entry}}^\star, z_{\text{exit}}^\star) =
    (2.15, 0.50),
\end{equation}
generating $9$ round-trip trades, an annualised net Sharpe ratio of $J(\theta^\star)=0.67$, and a net P\&L of $\$16{,}941$ on the initial capital of $\$100{,}000$.
Figure~\ref{fig: optimal equity curve 2018-2023} depicts the equity curve under optimal trading.

    \begin{figure*}[b]
    \centering
    \includegraphics[width=\textwidth]{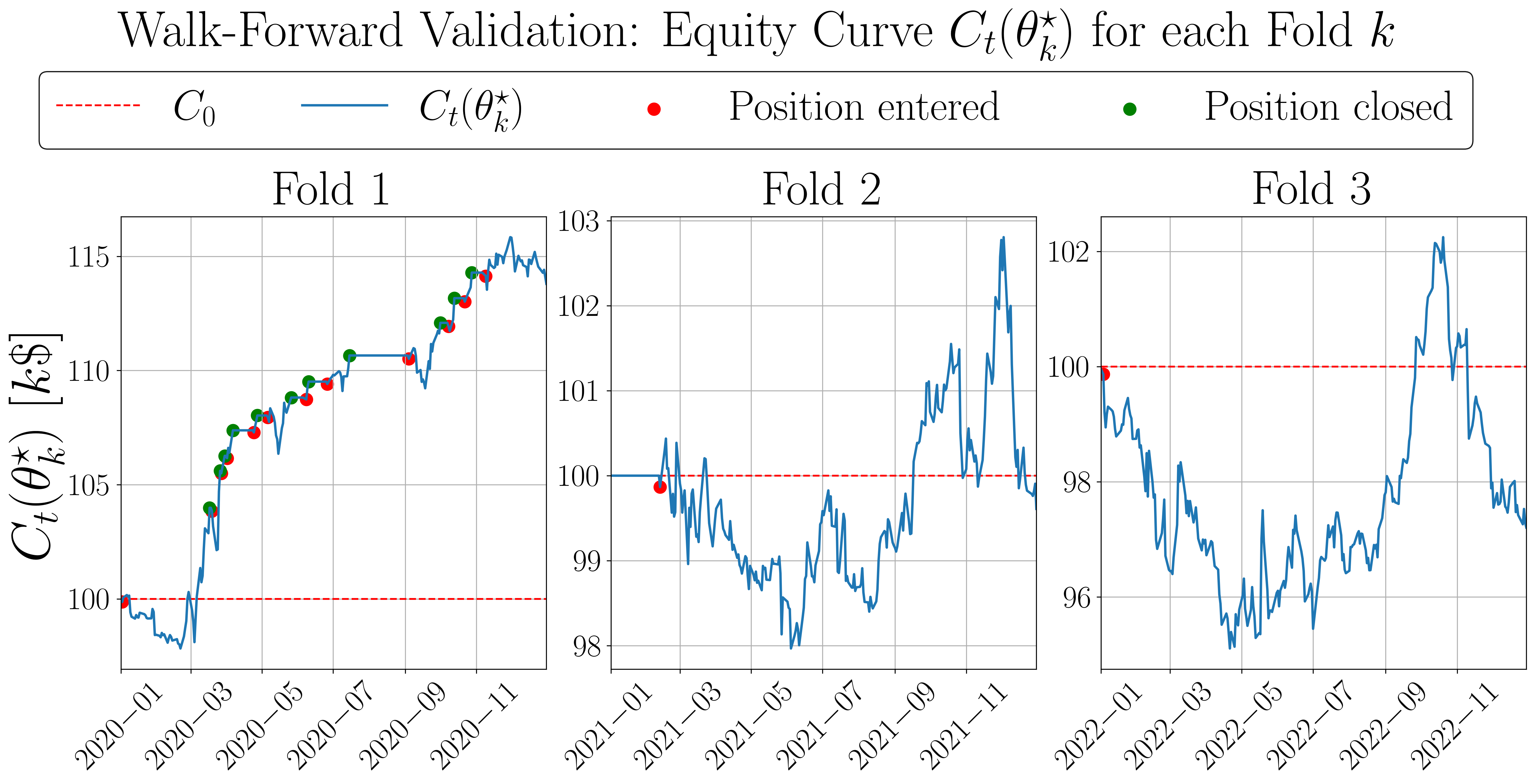}
    \caption{Walk-forward validation equity curve during the OOS window $\mathcal{S}_k$ for each fold $k=1,2,3$.}
    \label{fig: walk-forward validation equity curves}
\end{figure*}

\section{Robustness Analysis}
\label{section: robustness}

We first assess the robustness of the optimal policy $\theta^\star=(2.15,0.50)$ selected in Section~\ref{section: trading policy} to its auxiliary parameters. 
We then evaluate the stability of the optimisation procedure itself using walk-forward testing.

\subsection{Sensitivity Analysis}
\label{sec:sensitivity}

We vary each auxiliary parameter (transaction cost $c$, risk budget $w$, volatility-estimation window $L$, and maximum exposure $N_{\max}$) individually while holding the remaining parameters at their baseline values: $c=0.20\%$, $w=0.50\%$, $L=64$ days, $N_{\max}=25\%$. 
For each parameter set, the policy $\theta$ is re-optimised as in \eqref{eq: annualised net Sharpe ratio - objective function}.

\begin{table}[h] 
    \centering 
    \small 
    \begin{tabular}{lccc} 
        \toprule 
        & Range & $\theta^\star=(2.15,0.50)$ for & $J^\star$ (range) \\ 
        \midrule $c$ & $0.01$--$1.00\%$ & $0.05\%\le c\le0.50\%$ & $0.32$--$0.77$ \\ 
        $w$ & $0.10$--$1.00\%$ & $w\ge0.50\%$ & $0.55$--$0.72$ \\ 
        $L$ & $3$--$478$ days & $L\ge6$ days & $0.67$--$0.72$ \\ 
        $N_{\max}$ & $5$--$100\%$ & $N_{\max}\le25\%$ & $0.55$--$0.72$ \\ 
        \bottomrule 
    \end{tabular} 
    \caption{Sensitivity of $\theta^\star$ to auxiliary parameters.} 
    \label{tab:sensitivity} 
\end{table}

The baseline policy $(2.15,0.50)$ remains optimal across the majority of the parameter grid (see Table~\ref{tab:sensitivity}). 
At the boundaries of some grids, the optimum shifts toward lower entry and exit thresholds, corresponding to more frequent trading. 
This occurs at low transaction costs and risk budgets, shorter volatility windows, and higher exposure limits, where the penalty associated with increased trading activity or position size is reduced. 
The resulting variation in $J^\star$ is moderate across most parameter ranges. 
Overall, the recurrence of the same $\theta^\star$ across broad ranges of the auxiliary parameters provides evidence that the selected policy is not highly sensitive to their precise calibration.

\subsection{Walk-Forward Validation}
\label{section: walk-forward}

To assess whether the optimised policy generalises beyond the 2018-2023 sample used for calibration, we conduct a walk-forward analysis. 
For each of three folds, we re-optimise $\theta$ using a rolling two-year training window $\mathcal{T}_k$ and then apply the resulting policy, without further adjustment, to the following one-year OOS window $\mathcal{S}_k$. 
The folds are defined as $\mathcal{T}_1=[2018,2020)$ through $\mathcal{T}_3=[2020,2022)$, with each training window followed by its corresponding one-year test window. 
We then concatenate the three OOS return series and compute the aggregate OOS Sharpe ratio $J^{\text{oos}}$.

\begin{table}[h]
    \centering
    \small
    \begin{tabular}{cccccc}
        \toprule
        Fold $k$ & $z_{\text{entry}}^\star$ & $z_{\text{exit}}^\star$ & $J_{\mathcal{S}_k}^\star$ & Net P\&L & $n_{\text{trades}}$ \\
        \midrule
        $1$ & $1.30$ & $0.65$ & $2.14$  & $\$13{,}794$  & $12$ \\
        $2$ & $2.15$ & $0.35$ & $-0.07$ & $-\$391$    & $1$  \\
        $3$ & $2.00$ & $0.50$ & $-0.50$ & $-\$2{,}796$ & $1$  \\
        \bottomrule
    \end{tabular}
    \caption{Walk-forward OOS results by fold.}
    \label{tab:walk-forward}
\end{table}

The OOS results are highly uneven across folds (see Figure~\ref{fig: walk-forward validation equity curves}). 
Fold~1 generates most of the aggregate performance, with 12 trades and a Sharpe ratio of $2.14$, whereas folds~2 and~3 each generate only one trade and negative returns. 
The optimal thresholds also vary across folds, indicating that the calibrated policy is sensitive to the training period. 
Overall, the aggregate OOS Sharpe ratio $J^{OOS}=0.64$ provides evidence of positive performance outside the calibration sample, but the strong variation across folds suggests that this performance is not consistent across time.

\subsection{COVID-19 Dislocation and Regime-Dependence}
\label{sec:covid}

\begin{figure}[t]
    \centering
    \includegraphics[width=\textwidth]{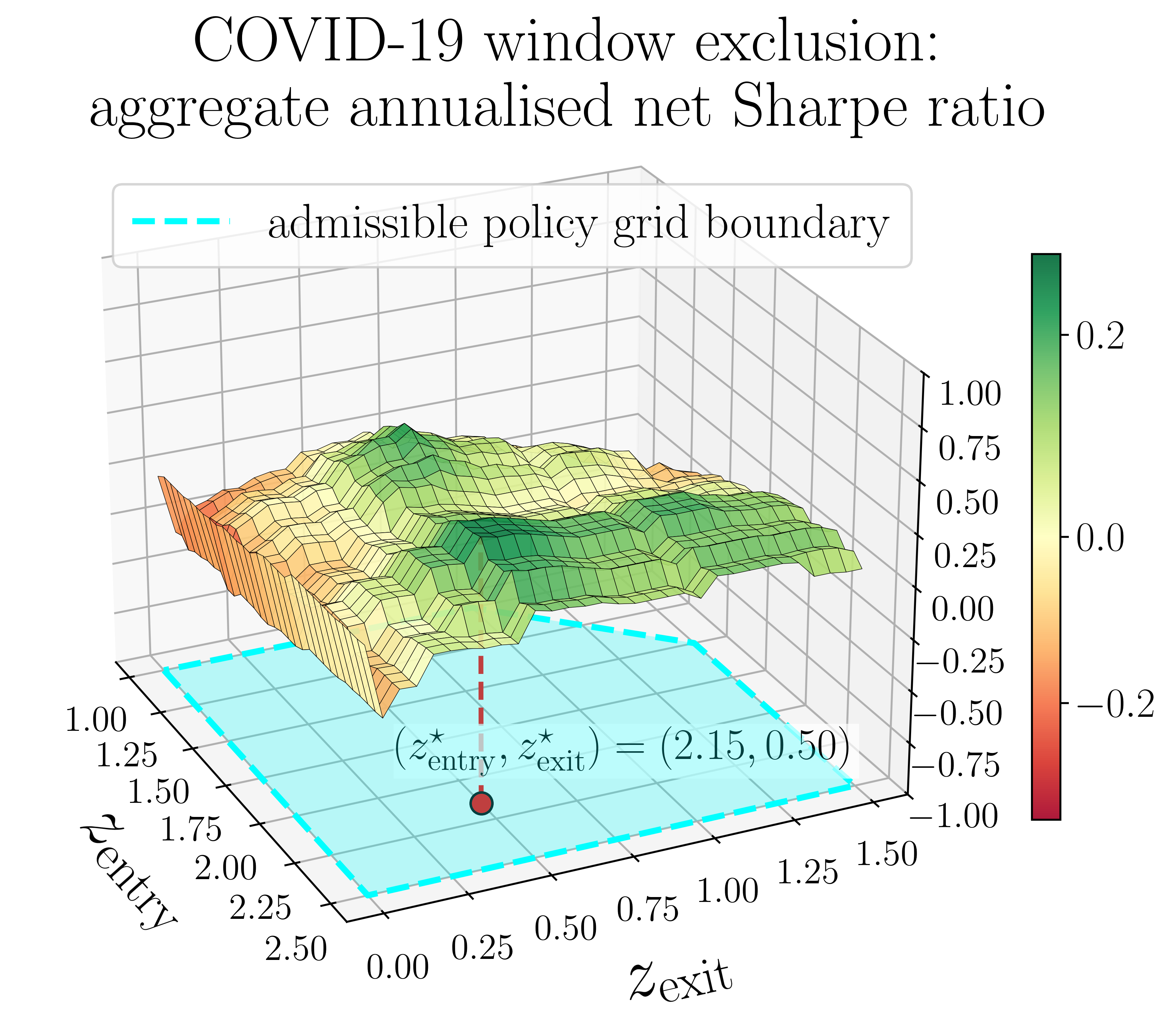}
    \caption{Aggregate annualised net Sharpe ratio surface over the admissible grid $\Theta$ during the Pre- and Post-COVID windows.}
    \label{fig: sharpe surface without covid 19}
\end{figure}

The pronounced asymmetry across walk-forward folds is closely associated with the COVID-19 volatility dislocation.
An inspection of the optimal equity curve over the full 2018--2023 sample shows that a substantial share of total net P\&L is generated during the February--April 2020 period, characterised by an unusually large and rapid increase in volatility followed by sharp mean reversions -- exactly the kind the strategy is designed to exploit.
The concentration of performance in this period therefore raises concerns about regime dependence and the extent to which the full-sample results generalise to more typical market conditions.
\indentation 
To isolate this, we re-run the analysis with the COVID window excluded, treating the pre- and post-COVID periods as two independent segments, each with its own rolling-volatility buffer and equity reset. 
This prevents the exclusion window from creating an artificial jump in returns or contaminating the rolling volatility estimates at the splice point.
\indentation
The optimiser still selects $\theta^\star=(2.15,0.50)$, but performance deteriorates materially; see Figure~\ref{fig: sharpe surface without covid 19} for the aggregate annualised net Sharpe ratio over the admissible grid.
Under optimal trading, the trade count falls from $9$ to $8$, the annualised Sharpe ratio from $0.67$ to $0.28$, and net P\&L from $\$16{,}941$ to $\$5{,}759$. 
With only $8$ trades, these results carry considerable sampling uncertainty, but the direction of the effect is unambiguous: a large share of the strategy's headline performance is attributable to the COVID dislocation, and its edge outside acute volatility regimes appears modest at best.

\subsection{Adjusted and Deflated Sharpe Ratios}
\label{section: asr-dsr}

The Sharpe ratios reported so far have two limitations. 
First, they assume Gaussian returns, whereas the daily return on capital is mostly zero with occasional large moves. 
Second, $\theta^\star$ was selected as the best-performing policy among $N=901$ candidates.
We therefore compute the Adjusted Sharpe Ratio (ASR), which accounts for skewness and excess kurtosis using the Pezier--White adjustment \cite{Pezier_White_ASR}, and the Deflated Sharpe Ratio (DSR), which accounts for the multiple-policy search by comparing the observed daily Sharpe ratio with the level expected from $N$ zero-skill strategies \cite{Bailey_LopezDePrado_DSR}.
Both measures are computed on the COVID-excluded sample.
\indentation 
The ASR is $0.281$, essentially unchanged from the raw Sharpe ratio of $0.28$, indicating that non-normality has little effect on the risk-adjusted performance estimate at this Sharpe level.
The DSR is $0.423$, implying that the observed Sharpe ratio is not statistically distinguishable from the maximum Sharpe ratio that could arise by chance when searching over $901$ zero-skill policies.
While this does not establish that the strategy has no genuine edge, it indicates that the in-sample optimisation provides limited statistical evidence of one, consistent with the weak performance observed after excluding the COVID-19 dislocation.

    \section{Out-of-Sample Performance}
\label{section: oos performance}

We evaluate $\theta^\star=(2.15,0.50)$ on the 2023--present period as an OOS test. 
The statistical parameters $\hat\beta,\hat\alpha,\hat m,\hat\sigma_u$ are frozen at the estimates obtained from the statistical analysis (2013--2018), while the policy parameters $c,w,N_{\max},L$ are frozen at their baseline values (see Section~\ref{section: trading policy}). 
At the trade frequency observed in-sample during the 2018-2023 window (roughly $2$--$3$ trades per year outside the COVID window), this window is expected to yield only $5$--$10$ trades, too few for any resulting Sharpe ratio to be treated as a reliable performance estimate.
\indentation 
The results of the OOS test are reported in Table~\ref{tab: oos performance 2023-present}.
The strategy enters a single position early in the window and force-liquidates it at the end, as the exit threshold $z_{\mathrm{exit}}$ is never crossed.
Equity declines almost monotonically from a few weeks after entry, resulting in a net P\&L of approximately $-\$28{,}000$ before liquidation.
With only one trade, the reported Sharpe ratio of $-1.19$ is not a meaningful estimate of OOS performance. 
However, the underlying failure mode is informative. 
Re-estimating the hedge ratio on the OOS window yields $\hat\beta_{\mathrm{OOS}} = -0.2376$.
Compared with the analysis window $\hat\beta = 1.6618$, this is not just a different magnitude but also the opposite sign.
A post-hoc Engle--Granger test on the resulting residuals, using the adjusted critical values, no longer rejects the null of no cointegration at the 5\% level ($p = 0.12$).
Since the residuals cannot be shown to be stationary, $\hat\beta_{\mathrm{OOS}}$ should not be interpreted as a meaningful long-run coefficient.
Although the test has limited power on a sample of this length, the result is consistent with the deterioration in OOS performance reflecting a breakdown of the cointegrating relationship between the two companies.
However, the single-trade sample does not allow a definitive attribution.

\begin{table}[h]
    \centering
    \small
    \begin{tabular}{ccc}
        \toprule
        $J^\star$ & Net P\&L & $n_{\text{trades}}$ \\
        \midrule
        $-1.19$ & $-\$28{,}131$ & $1$ \\
        \bottomrule
    \end{tabular}
    \caption{Out-of-sample test (2023--present) with frozen parameters.}
    \label{tab: oos performance 2023-present}
\end{table}

\subsection{Rolling Hedge Ratio}
\label{section: rolling hedge ratio}

The sign flip in $\hat\beta_{\text{OOS}}$ motivates a diagnostic test using a time-varying hedge ratio $\beta_t$. 
We consider two standard estimators: 
\ital{rolling OLS}, which re-estimates $(\alpha_t,\beta_t)$ over a trailing $M$-day window ($M=504$, approximately two trading years);
a \ital{Kalman filter} \cite{Kalman_Filter}, which models $\beta_t$ as a random walk,
\begin{equation*}
\beta_t=\beta_{t-1}+w_t, \qquad w_t \overset{\mathrm{i.i.d.}}{\sim} \mathcal{N}(0,\sigma_w^2),
\end{equation*}
and recursively updates $\hat\beta_{t\mid t}$ from the observation equation
\begin{equation*}
X_t^{\text{PEP}}=\alpha+\beta_t X_t^{\text{KO}}+\varepsilon_t.
\end{equation*}
In both adaptive variants, the intercept $\hat\alpha$ is re-estimated jointly with the hedge ratio, so that $(\hat\alpha_t,\hat\beta_t)$ evolve over time, and the spread $u_t = X_t^{\mathrm{PEP}} - \hat\alpha_t - \hat\beta_t X_t^{\mathrm{KO}}$ is standardised using its own trailing $M$-day statistics:
\begin{equation*}
    z_t = \frac{u_t - \mu_t}{\sigma_t},
\end{equation*}
where
\begin{equation*}
    \mu_t = \frac{1}{M}\sum_{s=t-M}^{t-1} u_s, \qquad
    \sigma_t^2 = \frac{1}{M-1}\sum_{s=t-M}^{t-1}\left(u_s - \mu_t\right)^2 .
\end{equation*}
All quantities use only information up to $t-1$ only, preserving the no-look-ahead property of the static backtest. 
Once a position is opened, the parameters $(\hat\alpha,\hat\beta,\mu,\sigma)$ are frozen at their entry values and used to evaluate the exit signal, so that entry and exit refer to the same spread definition.
However, since neither estimator is subjected to the same walk-forward, sensitivity, or ASR/DSR analysis as the static model, the results are interpreted as diagnostic rather than as evidence of a validated strategy.
A detailed account of the implementation can be found in the extended version of this paper \cite{graziano_pep_ko_pairs_trading}.

\begin{figure}[h]
    \centering
    \includegraphics[width=\textwidth]{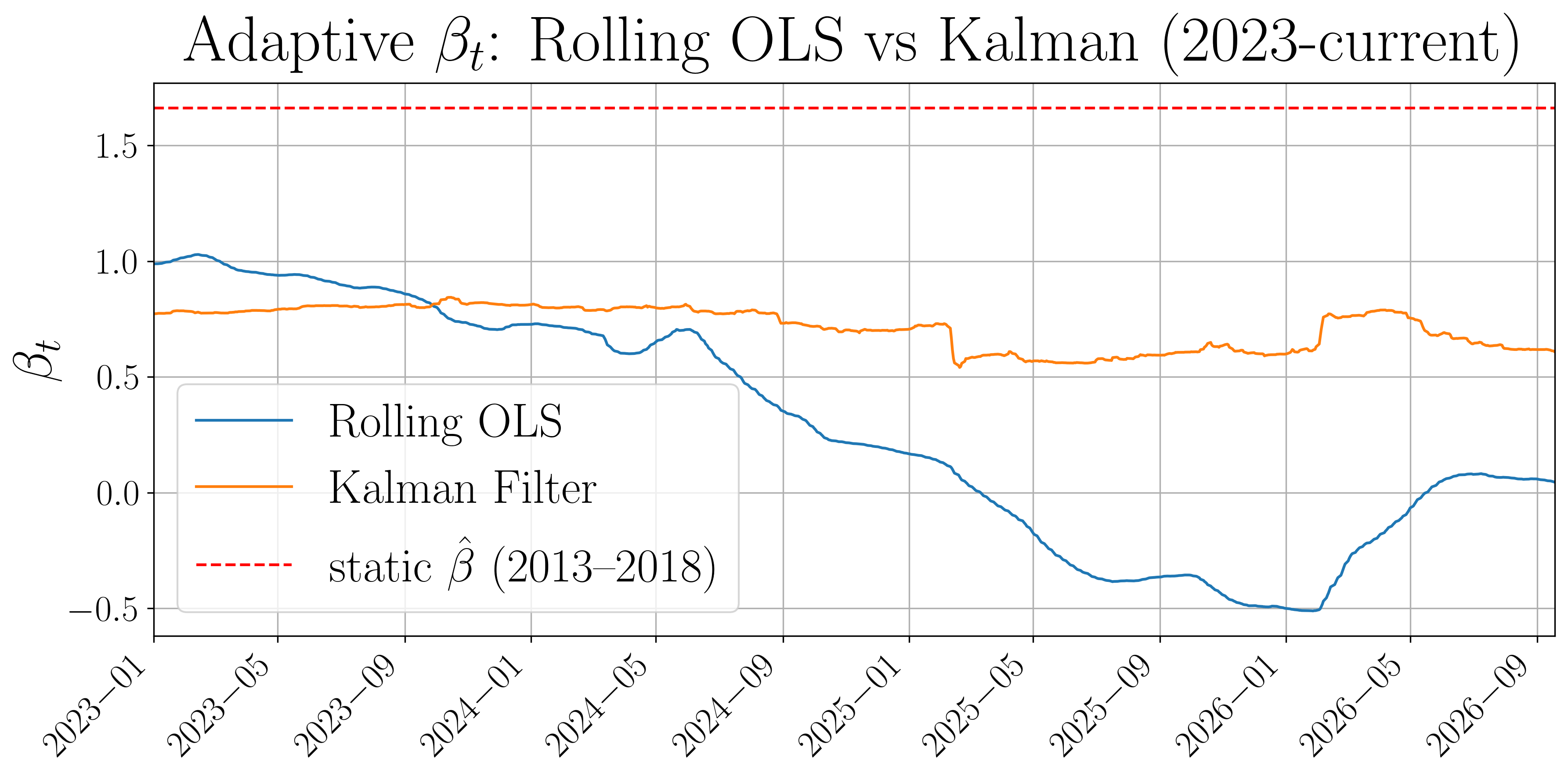}
    \caption{Time-varying hedge ratio $\beta_t$, 2023--present: rolling OLS vs. Kalman filter.}
    \label{fig: rolling hedge ratio}
\end{figure}

\begin{table}[h]
    \centering
    \small
    \begin{tabular}{lccc}
        \toprule
        Estimator & $J^\star$ & Net P\&L & $n_{\text{trades}}$ \\
        \midrule
        Rolling OLS ($M=504$) & $-1.05$ & $-\$27,163$ & $1$ \\
        Kalman filter          & $-0.88$ & $-\$32,752$ & $5$ \\
        \bottomrule
    \end{tabular}
    \caption{OOS test (2023--present) with adaptive hedge ratio and static policy $\theta^\star=(2.15,0.50)$.}
    \label{tab: rolling hedge ratio OOS performance}
\end{table}

As shown in Figure~\ref{fig: rolling hedge ratio}, rolling OLS captures the hedge ratio drift identified earlier: $\beta_t$ decreases from approximately $1$ to $-0.5$ before partially recovering, consistent with the sign change found in the earlier OOS re-estimation.
The strategy remains flat for the first 18 months and then executes a single trade in mid-2024, which is closed only by forced liquidation at the end of the sample.
The result is a net loss of approximately $\$27{,}000$ (see Table~\ref{tab: rolling hedge ratio OOS performance}), similar to the static backtest.
The Kalman filter produces instead a much smoother $\beta_t$ trajectory that does not exhibit the sign change. 
It executes five trades before forced liquidation and ends the sample with a net loss of approximately $\$33{,}000$, larger than that of the static model.
Allowing the hedge ratio to adapt therefore does not improve OOS performance: rolling OLS reproduces the static loss, and the Kalman filter worsens it, despite the adaptive estimators tracking the changing PEP/KO relationship.
With only $n=1$ and $n=5$ trades, neither Sharpe ratio is statistically meaningful.
\indentation 
These results provide evidence, although limited by the small number of trades, that hedge-ratio drift alone does not account for the failure of the static model over the 2023--present window.
Even when $(\hat\alpha_t,\hat\beta_t)$ and the standardisation parameters are re-estimated using only past data, the resulting spread does not revert to equilibrium quickly enough for the exit rule to be triggered before the end of the sample.
This is consistent with the post-hoc Engle--Granger result above, and suggests a change in the relationship between PEP and KO, such as a spread that no longer exhibits the mean reversion assumed by the policy.

    \section{Conclusion}
This paper examined whether a cointegration-based PEP/KO pairs trading strategy remains statistically robust and economically exploitable once transaction costs, model uncertainty, and policy selection are taken into account. 
The evidence presented in this paper provides limited support for this proposition.
\indentation 
The 2013--2018 analysis in Section~\ref{section: statistical analysis} identifies a statistically significant cointegrating relationship between PEP and KO, with a slowly mean-reverting spread (half-life $\approx 32$ days), providing a basis for the policy developed over the 2018--2023 optimisation period in Section~\ref{section: trading policy}. 
Section~\ref{section: robustness} shows that, although the resulting policy is relatively robust to its auxiliary parameters, walk-forward validation exposes that its performance is concentrated in a single fold; excluding the COVID-19 market dislocation reduces the Sharpe ratio by more than half. 
After accounting for the 901 policies considered, a Deflated Sharpe Ratio of 0.423 further indicates that the in-sample results provide limited evidence of genuine skill.
\indentation 
The 2023--present out-of-sample test in Section~5 provides further evidence against the robustness of the strategy: it produces a single losing trade with a substantial loss, while the estimated hedge ratio changes markedly and reverses sign relative to its 2013--2018 estimate. 
A post-hoc Engle--Granger test no longer rejects the null of no cointegration ($p=0.120$).
Moreover, the rolling OLS and Kalman filter adaptive estimation of the hedge ratio fail to restore a mean-reverting spread: rolling OLS reproduces the static loss, while the Kalman filter results in an even larger one. 
Together, these results are consistent with a regime change in the PEP/KO relationship, although the small number of trades and the limited power of the test prevent a definitive conclusion.
\indentation 
More broadly, the results illustrate that a statistically significant and economically plausible cointegrating relationship within one estimation regime does not imply that its trading dynamics will remain stable out of sample. 
For PEP and KO, the mean-reversion characteristics observed during the estimation and optimisation periods did not persist into 2023--present, and the concentration of in-sample performance around the COVID-19 volatility regime further limits the evidence for a durable trading opportunity. 
Real-time regime-adaptive hedging is a natural direction for further investigation, but the pair's low trading frequency would continue to limit statistical power and make reliable validation challenging.

    \printbibliography

\end{document}